\documentstyle[aps,epsf,preprint]{revtex}
\tighten

\begin{document}
\input epsf
\draft

\title{Quiet and Noisy Metastable Voltage States in High-T$_c$ Superconductors}
\author{G. Jung$^*$}
\address{Department of Physics, Ben Gurion University of the Negev,\\
P.O. BOX 653, 84105 Beer-Sheva, Israel}
\author{B.Savo}
\address{ Dipartimento di Fisica and INFM, Universit{\'a} di Salerno,
84081 Baronissi (SA), Italy}
\author{Y.Yuzhelevski}
\address{Department of Physics, Ben Gurion University of the Negev,\\
P.O. BOX 653, 84105 Beer-Sheva, Israel}
\maketitle

\begin{abstract}
The interaction between the telegraph noise and background voltage
fluctuations in the current induced dissipative state of high-T$_c$
BiSrCaCuO thin films has been investigated. Experimental time records of the
voltage drop across current biased thin film strips show markedly different
background noise traces in the up and down telegraph states. Detailed
analysis demonstrates that fluctuations around the telegraph voltage levels
are due to a unique background noise process. The apparent quiet and noisy
voltage states are due only to differences in the effective frequency
bandwidth at which background noise is seen at distinct telegraph levels.
Changes of the  background noise variance ratio with changing bias current
follow changes of the statistical average lifetimes of the random telegraph
process.
\end{abstract}

\section{Introduction}

Voltage noise in the thermodynamically superconducting state is associated
with dissipation induced by motion of magnetic flux structures and/or action
of intrinsic Josephson junctions. In good quality high-T$_c$ superconducting
(HTSC) samples the dissipation caused by the flow of transport currents
along the superconducting planes is dominated by dissipative flux processes.
Flux noise due to randomness in vortex matter dynamics converts into
observable voltage fluctuations by means of an intrinsic flux-to-voltage
conversion mechanism \cite{clem,ens,jap1f}. Low frequency flux and voltage
noise in HTSC systems typically appears as wide band Gaussian fluctuations
with a $1/f$-like power spectral density (PSD). $1/f$-like noise is
frequently accompanied by characteristic non-Gaussian random telegraph noise
(RTN) components. Telegraph signals in HTSC systems were detected in
magnetic flux noise at low \cite{clarke,clarkecorrel,uk} and high magnetic
fields \cite{argonne}, magnetically modulated microwave absorption
\cite{mmm} and in voltages appearing across dc current biased thin films
\cite{jap1f,jap,epl,isra,kiss,prb}.

In the simplest case of a two-level random telegraph signal (dichotomous
noise) the observable switches randomly between two fixed levels, referred
to as "up" and "down" level. Generation of dichotomous noise can generally
be traced to an action of a two-level fluctuator (TLF) consisting of two
energy wells separated by a barrier. The system undergoes thermally
activated or tunnel transitions between the wells corresponding to random
switching of the measured observable, for a review see ref. 4. In reality,
random telegraph signals deviate from the ideal two-level fluctuator
picture. First of all, experimentally observed RTN always appears on the
background of noise contributed by other random processes in the sample and
by instruments in the electronics chain. Waveforms of experimentally
observed HTSC random telegraph signals frequently exhibit exotic features,
such as  multi--level switching \cite{clarke,uk,toledo} and modulation of
telegraph amplitude and/or switching frequency by yet another telegraph
signal \cite{eucas,upon}.

Among many exotic manifestations of RTN signals in HTSC systems, events
demonstrating different traces of background noise at different RTN levels,
or in other words, background noise which changes synchronously with the
telegraph signal, deserves particular attention. This phenomenon was first
observed in flux noise experiments performed at zero field cooled HTSC
samples and since then is referred to as "noisy and quiet metastable states"
\cite{clarke}. Pronounced asymmetry in the background noise variance at
distinct telegraph levels has subsequently been observed by us in the
voltage noise of current biased HTSC thin films \cite{eucas}. The appearance
of quiet and noisy telegraph states was tentatively interpreted as a
signature of vortex hopping from a site where it is relatively mobile to a
site where it is much more restricted spatially \cite{clarke}. This
assumption imposes strong conditions on the model of an active two-level
fluctuator by requiring the TLF energy wells to have different curvature.
Consequently, the attempt frequencies for two distinct wells must also be
assumed to be different. Thus a TLF responsible for the appearance of quiet
and noisy metastable states deviates markedly from the classical TLF
scenario of two symmetric wells separated by an asymmetric barrier
\cite{koganbook,kirton}.

It is worth remembering here that similar exotic random telegraph waveforms
have been observed in non-superconducting solid state systems
\cite{wakai,kirton,farmer,others}. The non-superconducting quiet and noisy
RTN events were generally ascribed to interactions between localized
structural defects and active two-level fluctuators in small size systems.
Nevertheless, mechanisms responsible for synchronous switching of RTN level
and background noise intensity as well as modulation of the telegraph
waveforms in HTSC samples cannot be consistently explained by evoking
similar defect-fluctuator interactions \cite{clarke}.

This paper is devoted to the experimental investigation of the nature of
quiet and noisy metastable states appearing in random telegraph voltage
noise in zero field cooled BiSrCaCuO thin film strips which are driven into
a dissipative state by bias current flow.

\section{Experimental}

The experiments were performed with 300 nm thick, c-axis oriented high
quality thin  BiSrCaCuO 2212 films fabricated by means of molecular beam
epitaxy. The details of sample preparation and characterization are reported
elsewhere \cite{mbe}. The films had a very high residual resistance ratio,
R$_{300/100}\geq 3.3$, T$_c(R=0)$ above 86 K, and critical current density
$J_c(4.2 K)\sim 10^5 A/cm^2$. X--ray diffraction spectra of the films showed
only peaks of the pure 2212 phase and strong preferential orientation with
c--axis perpendicular to the substrate plane. The films were patterned into
50 $\mu$m wide strips with large silver covered contacts pads on both ends
of the strip and voltage pick--up leads separated by 50 $\mu$m. In the
experiments the voltage signal developed under $dc$ current flow in the
strip was delivered to the top of cryostat, amplified by a low noise
preamplifier, and processed by a signal analyzer. Several random telegraph
events were detected in many, relatively narrow, noisy window ranges of
temperature, current flow and associated magnetic fields. In this paper we
concentrate however only on RTN events observed in the current induced
dissipative state at zero applied magnetic field.

An example of a  RTN waveform appearing in one of our strips at 77 K is
shown in Fig. \ref{rtn-record}. Even a brief examination of the experimental
record convinces one that the apparent background noise intensities at
distinct telegraph levels are markedly different. This is a clear
manifestation of the quiet and noisy metastable states seen in the form of
voltage fluctuations. It should be emphasized that the appearance of quiet
and noisy metastable states in current biased HTSC samples is not restricted
to a particular type of a sample or deposition technique. We have previously
reported similar events in BSCCO films obtained by liquid phase epitaxy on
NdGaO$_3$ substrates \cite{eucas}.

The switching in the intensity of the background noise synchronous with the
RTN signal suggests that some form of statistical correlations between RTN
and background fluctuations may exist. To get a deeper insight into this
puzzling phenomenon we have performed a detailed statistical analysis of the
RTN waveforms and background fluctuations. The experimentally observed
background noise has been investigated by analyzing Gaussian distributions
of voltage fluctuations around mean voltages of the RTN levels. For each
current flow the analysis was performed by averaging the results of at least
5 time records sampled in 40960 points. The RTN components of the
experimental record have been initially analyzed by annotating the time
instances at which the system undergoes transitions between RTN states,
determining the time lengths of individual pulses and heights of individual
RTN amplitudes, building their histograms, and finding the statistical
average values. This procedure is straightforward for clean RTS signals,
well above the background noise intensities. However, this approach fails
completely in the case of "noisy" records, for which it becomes difficult to
determine the precise moment at which transitions occur. To enable analysis
of the experimental records strongly perturbed by the background noise we
have developed a new procedure of RTN analysis in the time domain which is
based on differences between statistical properties of the Gaussian
background noise and Marcovian RTN fluctuations (Section \ref{tda}.

Statistical analysis confirmed that variances of background fluctuations
around up and down telegraph levels are indeed different. Moreover, we have
found that the ratio between variances $\sigma_{up}/\sigma_{dn}$ changes
markedly with changing current flow, as shown in Fig. \ref{variance-ratio}.
The dependence of the variance ratio on bias current should not be
surprising, since all statistical average parameters of the RTN fluctuations
in HTSC systems are known to change with changing current flow
\cite{jap1f,jap,epl,isra,prb}. The evolution of RTN lifetimes with changing
bias current as determined from experimental records is illustrated in Fig.
\ref{tempi}, while the current dependence of RTN amplitude is plotted in
Fig. \ref{variance-ratio} together with the variance ratio.

Fig. \ref{variance-ratio} shows that  current induced changes of the
variance ratio follow closely the RTN amplitude evolution. This feature may
be evoked as an argument in the favor of possible statistical correlation
between RTN and background fluctuations. However, a careful examination of
the variance ratio behaviour reveals that at a certain current flow, the
variances of the background noise at RTN levels become equal,
$\sigma_{up}=\sigma_{dn}$. Moreover, one finds that with further current
increase the noisy and quiet metastable states are interchanged. The current
flow at which the variances become equal corresponds to the symmetrizing
current $I_s$ at which $\tau_{up}=\tau_{dn}$ and the RTN waveform is
symmetric: compare Figs. \ref{tempi} and \ref{variance-ratio}.

The disappearance of  differences between the background fluctuations around
RTN levels at the symmetrizing current strongly suggests that the apparent
differences in RTN background noise variances are associated with
differences in RTN lifetimes at distinct metastable levels. This translates
directly into differences in the effective bandwidth in which the background
noise at each level is observed in the experiment. In the time domain the
mean square noise around each RTN level is
\begin{equation}
<\sigma_{up(dn)}^2>
=\frac{1}{T}\int_{0}^{T}{[U(t)-\overline{U}_{up(dn)}]^2dt},
\end{equation}
where $U(t)$ is the departure from the mean value $\overline{U}_{up(dn)}$ of
the signal, and $T$ is a time interval. On the other hand, in the frequency
domain, the variance of a signal with a zero mean can be expressed as
\begin{equation}
<\sigma^2> =\int_{f_{min}}^{f_{max}}{S(f)df}, \label{varint}
\end{equation}
where the bracket denotes ensemble averaging, $S(f)$ is the spectral density
function (PSD). The maximum frequency in the record and the upper limit for
the integral (\ref{varint}) is  set by the data sampling frequency,
$f_{max}=1/\Delta t$, where $\Delta t$ is the time interval between data
points. The lower frequency limit of the experimental bandwidth is set by
the inverse of the average RTN lifetime, $f_{min}=1/\tau$
\cite{kleinpenning1,strassila,kleinpenning2}. To compare the real
intensities of the background noise at distinct RTN level and to establish
whether quiet and noisy metastable states exist in physical reality one
should  calculate the variance ratio directly from (\ref{varint}). This
requires the functional form of $S(f_{bckgnd})$ for the background noise.

One can determine $S(f)_{bckgnd}$, assuming the RTN and background
fluctuations are uncorrelated, by subtracting $S(f)_{rtn}$ from
$S(f)_{exp}$. For a pure two-level RTN signal \cite{machlup},
\begin{equation}
S(f)_{rtn}=4\Delta V^2{{(\tau_{up}
\tau_{dn})^2}\over{(\tau_{up}+\tau_{dn})^3}} ~{{1}\over{1+4\pi^2 f^2
/f_c^2}}\label{machlup}
\end{equation}
where $f_c=\tau^{-1}_{up}+\tau^{-1}_{dn}$. The spectrum of a pure RTN
contribution can be calculated from (\ref{machlup}), by inserting the
amplitude $\Delta V$, and average lifetimes $\tau_{up}$, and $\tau_{dn}$
obtained from statistical analysis of the experimental time records.
However, this approach cannot be applied to signals in which RTN and
background fluctuations are correlated. In this case the spectrum
$S(f)_{exp}$ also contains  unknown cross-correlation term,
$S(f)_{exp}=S(f)_{rtn}+S(f)_{bckgnd}+S(f)_{rtn,bckgnd}$. In our experiments
we find strong indications that RTN and background noise fluctuations may be
correlated. Moreover, if quiet and noisy metastable states really exist it
is quite plausible that the background noise at distinct RTN levels may be
characterized by spectral densities not only with different intensities but
also with different functional forms of $S(f)$. Thus to investigate RTN
background noise problem one should create artificial time records
representing background fluctuations at each RTN level and find their
Fourier transforms. Noise records of the separate RTN levels can be obtained
by redistributing the experimental record into two subsets, each containing
only data points belonging to two given RTN state, as described in the next
section.

\section{RTN analysis in time domain}\label{tda}

The proposed analysis procedures are based on the assumption that the
telegraph noise constitutes a discrete Marcovian dichotomous signal with
Poisson statistics of the lifetime distribution and a single amplitude
$\Delta V$, while the fluctuations within telegraph levels are due only to
the background noise which is assumed to be Gaussian.

In the experiment, the continuous signal $U(t)$ is sampled with a frequency
$f_c=1/\Delta t$ into a digital record $\{U_n\}$ of  the length $N\Delta t$,
containing $N$ data points equally spaced in chronological order, $n = (1,
2, ...N)$. First, we fit amplitude histogram $\{U_n\}$ to the two-Gaussian
distribution $G(U_n)$ corresponding to the sum of background noise
distributions around up and down RTN levels:
\begin{equation}
G(U_n)=G_{dn}(U_n)+G_{up}(U_n)=\frac{A_{dn}}{\sigma_{dn}\sqrt{2\pi}}
e^{-\frac{(U_n-\overline{U}_{dn})^2}{2\sigma_{dn}^2}}+
\frac{A_{up}}{\sigma_{up} \sqrt{2\pi}}
e^{-\frac{(U_n-\overline{U}_{up})^2}{2\sigma_{up}^2}},
\label{Gauss}
\end{equation}
where $\sigma_{dn}$ and $\sigma_{up}$ are the variances of the background
noise around the mean values $\overline{U}_{dn}$ and $\overline{U}_{up}$,
respectively, $A_{up}=N_{up}\Delta U$ and $A_{dn}=N_{dn}\Delta U$ are the
areas under the Gaussian curves, $\Delta U$ is the size of the amplitude
histogram bin, and $N_{up}$ and $N_{dn}$ stand for the total number of data
points in the record belonging to the respective $\{up\}$ and $\{dn\}$
telegraph state. The probability that a data point from $\{up(dn)\}$ takes a
value $[U_n-\frac{\Delta U}{2}, U_n +\frac{\Delta U}{2}]$ is
$G_{up(dn)}(U_n) \frac{\Delta U}{A_{up(dn)}}$ . We start from a point $n-1$
that with the probability ${\cal P}=1$ belongs to $\{dn\}$, i.e.,
$U_{n-1}\leq U^{up}_{min}$ (see Fig. \ref{gauss-fig}). The probability that
the next data point $n$, takes a value $U \epsilon[U_n-\frac{\Delta U}{2},
U_n +\frac{\Delta U}{2}]$ is given by a sum of the probabilities of two
alternative events, ${\cal P}^{up}_{n_{\{dn\}}}+{\cal P}^{dn}_{n_{\{dn\}}}$,
in which point $n$ belongs to the state $\{dn\}$ or $\{up\}$. The first term
is the product of the probabilities that the point $n$ takes a value from
the range $[U_n-\frac{\Delta U}{2}, U_n +\frac{\Delta U}{2}]$ and that a
transition from $\{dn\}$ to $\{up\}$ has occurred in the time between
acquisition of the data points $n-1$ and $n$,
\begin{equation}
{\cal P}^{up}_{n_{\{dn\}}}= G_{up}(U_n) \frac{\Delta U}{A_{up}} \frac{\Delta
t}{\tau_{dn}} .\label{nisup}
\end{equation}
The second term describes the joint probability of an event in which point
$n$ takes the same value and no transition occurs between data points $n-1$
and $n$,
\begin{equation}
{\cal P}^{dn}_{n_{\{dn\}}}= G_{dn}(U_n) \frac{\Delta U}{A_{dn}}
(1-\frac{\Delta t}{\tau_{dn}}) . \label{nisdn}
\end{equation}
A criterion for ascribing the data point $n$ either to the state $\{up\}$ or
to the state $\{dn\}$ can be based on comparing the probabilities
(\ref{nisup}) and (\ref{nisdn}). We define
\begin{equation}
{\cal F}_{dn}(U_n)=\frac{{\cal P}^{up}_{n_{\{dn\}}}}{{\cal
P}^{dn}_{n_{\{dn\}}}}=\frac{G_{up}(U_n)}{G_{dn}(U_n)}\frac{ A_{dn}\Delta
t}{A_{up}(\tau_{dn}-\Delta t)} . \label{critdn}
\end{equation}
Clearly, when ${\cal F}_{dn}>1$ then $n\; \epsilon \{up\}$, whereas for
${\cal F}_{dn}<1$  $n\; \epsilon \{dn\}$. In an analogous way, assuming that
point $n-1$ belongs to the state $\{up\}$, we can formulate the probability
of ascribing the data point $n$ to either $\{up\}$ or $\{dn\}$,
\begin{equation}
{\cal F}_{up}(U_n)=\frac{{\cal P}^{dn}_{n_{\{up\}}}}{{\cal
P}^{up}_{n_{\{up\}}}}= \frac{G_{dn}(U_n)}{G_{up}(U_n)}\frac{
A_{up}\Delta t}{A_{dn}(\tau_{up}-\Delta t)}.
\label{critup}
\end{equation}
For ${\cal F}_{up}>1$ point $n\;\epsilon \{dn\}$, whereas for ${\cal
F}_{up}<1$ we have $n\; \epsilon \{up\}$. For practical convenience one may
convert the probability criterion into the voltage criterion by solving
equations ${\cal F}_{dn}=1$ and ${\cal F}_{up}=1$. The solutions, $U^*_{dn}$
and $U^*_{up}$, respectively, determine the threshold voltages (see Fig.
\ref{gauss-fig}) which can be employed for fast redistribution of the
acquired data points between the RTN states. If the previously analyzed
point has been attributed to the state $\{dn\}$ then the next data point
will be ascribed to the same state if $U_n \leq U^*_{dn}$, and to the
opposite state otherwise. If a data point $n-1$ was ascribed to the state
$\{up\}$ then the point $n$ will belong to the same state when $U_n \geq
U^*_{up}$.

To determine the threshold voltages one needs to know $A_{up}$ and $A_{dn}$,
$\overline{U}_{up}$ and $\overline{U}_{dn}$, $\sigma_{up}$ and
$\sigma_{dn}$, and the average lifetimes  $\tau_{up}$ and $\tau_{dn}$. All
but $\tau_{up}$ and $\tau_{dn}$ are already known from the initial fit of
the experimental amplitude histograms to a two-Gaussian distribution
(\ref{Gauss}). The missing average RTN lifetimes can be determined through a
conventional statistical analysis of lifetime distributions or,
alternatively, evaluated from the areas under the relevant Gaussian curve,
provided that the total number of transition in the record, $k$, is known.
In a large $k$ approximation, $k\gg1$, the average RTN lifetimes can be
approximated by
\begin{equation}
\tau_{up(dn)}=\frac{2}{k} \frac{A_{up(dn)}}{\Delta U} \Delta t .
\label{tau}
\end{equation}
The total number of transitions in the experimental record is usually large
and the approximation $k\gg1$ is generally  well justified, nevertheless,
the total number of transitions in the record is still not known. The
missing $k$ value can be determined by tentatively redistributing data
points according to a simplified rough criterion, which does not require the
knowledge of average lifetimes, and subsequently performing iterative
fitting procedures of thus re-distributed records to the original
experimental Gaussian distributions.

In the approximation $\sigma_{up}=\sigma_{dn} =\sigma$. Eqs (\ref{critup})
and (\ref{critdn}) have the following solutions:
\begin{equation}
U^*_{dn}=\frac{\overline{U}_{dn}+\overline{U}_{up}}{2}+\frac{\sigma^2}{\Delta
U}\ln(\frac{\tau_{dn}}{\Delta t}-1)=\overline{U}_{dn}+\delta_{dn},
\end{equation}
\begin{equation}
U^*_{up}=
\frac{\overline{U}_{dn}+\overline{U}_{up}}{2}-\frac{\sigma^2}{\Delta
V}\ln(\frac{\tau_{up}}{\Delta t}-1)=\overline{U}_{up}-\delta_{up}.
\end{equation}
A rough criterion for the zero-order redistribution of data points can be
established by setting the logarithmic terms to unity. The resulting
approximate, overestimated voltage criteria read
\begin{equation}
\widetilde{U}^*_{dn}=\overline{U}_{dn}+\frac{\Delta
V}{2}+\frac{\sigma^2}{\Delta U},
\end{equation}
\begin{equation}
\widetilde{U}^*_{up}=\overline{U}_{up}+ \frac{\Delta
V}{2}-\frac{\sigma^2}{\Delta U}.
\end{equation}
After an initial tentative redistribution of data points from the
experimental record one can build an artificial time record of a pure RTN
and easily count the number of telegraph transitions $k_0$. Using $k_0$ as a
zero order approximation of the real $k$ one calculates $\tau_{up}$ and
$\tau_{dn}$ using (\ref{tau}). Next, the separation procedure is again
performed, this time with the help of the exact criteria (\ref{critup}) and
(\ref{critdn}), whichever is appropriate. Subsequently, the new, corrected
pure RTN record is created and the new number of transitions, $k_1$, is
counted. The iterations are repeated $i$ times, until the final number of
transitions $k_i=k$ is obtained. The final $k_i$ is the value for which
Gaussian distributions around each RTN state, as obtained through the
separation procedures, fit to the original experimental distributions
$G_{up}(U)$ and $G_{dn}(U)$. The described procedure converges rapidly and
typically $i<5$ iterations are needed to obtain a stable $k$ and calculate
the values of $\tau_{up}$ and $\tau_{dn}$.

The proposed procedure works well even for relatively noisy RTN signals, as
illustrated in Fig. \ref{example} showing a "noisy" RTN signal together with
a record of clean telegraph contribution revealed by means of the above
described procedure. Note that all telegraph jumps, even those strongly
perturbed by the background noise, can be seen in the pure RTN record. The
pure RTN record can be also employed as a guide to extract artificial
separate records of background noise at each RTN level from the data.

\section{Results and discussion}

The knowledge of the RTN average lifetimes, see Fig. \ref{tempi}, determined
by statistical analysis of the time record, and the acquisition rate allowed
us to determine the effective bandwidth for the background noise. Now, we
have to determine the functional form of the background noise spectra. For
that purpose, using the procedures described above, we created an artificial
time record of a pure RTN contribution and a record containing only the
background noise components. The background noise record was obtained by
subtracting, in the time domain, the pure telegraph contribution from the
total experimental record. Subsequently, for each current, we calculated the
spectral densities of background noise records.  An example of such analysis
for a current close to $I_s$ is shown in Fig. \ref{PSD}. The pure RTN
contribution has a Lorentzian shape described by Eq. (\ref{machlup}), while
the background noise power spectrum follows a $1/f^{0.5}$ power law within
the experimental frequency range. We have verified by further separation of
the background noise time record into two separate contributions
corresponding to the $\{up\}$ and $\{down\}$ RTN states that all the
background noise components are characterized by the same $1/f^{0.5}$ PSD.
Moreover, this functional form of the PSD does not change with changing
current within the entire noisy window range.

By inserting $S(f,I)=C(I)/f^{0.5}$ into Eq. (\ref{varint}) we obtain the
background noise variance at a given RTN level,
\begin{equation}
<\sigma^2> = \int_{\frac{1}{\tau(I)}}^{\frac{1}{\Delta t}}{S(f,I)df}=
\int_{\frac{1}{\tau(I)}}^{\frac{1}{\Delta t}}{\frac{C(I)}{f^{0.5}}df}=C(I)
\left[1-\left(\frac{2\Delta t}{\tau(I)}\right)^{0.5}\right],
\label{variance}
\end{equation}
where $C(I)$ is a current dependent constant characterizing the noise
intensity at a unit frequency and $\tau$ is the average lifetime at the
considered RTN level. We proceed by calculating, for each current, the ratio
between variances at two RTN levels using Eq. (\ref{variance}) with
$\tau(I)$ values determined from the statistical analysis of the pure RTN
waveform:
\begin{equation}
\frac{<\sigma^2_{dn}>}{<\sigma^2_{up}>} =\frac{C_{dn}(I)}{C_{up}(I)}\frac{
\left[1-\left(\frac{2\Delta t}{\tau_{dn}(I)}\right)^{0.5}\right]}{
\left[1-\left(\frac{2\Delta t}{\tau_{up}(I)}\right)^{0.5}\right]}.
\label{ratio}
\end{equation}

At this point we arrive at the crucial question about the real intensities
of the background noise around RTN levels, $C_{up}$ and $C_{dn}$. Let us,
for a moment, assume that the noise intensities on both levels are equal
$C_{up}/C_{dn}=1$. The variance ratio calculated from Eq. (\ref{ratio})
under the assumption of identical background noise intensities on both RTN
levels, is compared with the experimental variance ratio determined from the
width of the Gaussian background noise distributions in Fig. \ref{final}.
The agreement between the calculated and experimental results is excellent,
indicating that the background noise intensities on both RTN levels are
indeed identical. Therefore, we conclude that the difference in the
background noise traces, appearing as quiet and noisy metastable states, is
not due to the exotic structure of a two-level fluctuator but results from
the bandwidth limits imposed on the observable background noise by the
telegraph fluctuations.

It is worth emphasizing that it is the RTN mean lifetime that determines the
experimental bandwidth, and consequently noise variance at a given level,
and not the individual pulse duration. This feature is clearly seen in Fig.
\ref{last} showing a fragment of a time record demonstrating the effect of
quiet and noisy RTN states. For this record $\tau_{up}>\tau_{dn}$, and
voltage fluctuations around the up level appear to be much stronger than
those around the down level. Our claim that the difference in background
noise at two RTN levels is due to a different bandwidth in which we observe
fluctuations around a given state translates into the length of the time
window in which a given state is observed. One may then ask why are the
fluctuations within the pulse labeled as "1" in Fig. \ref{last} smaller then
the fluctuations of the pulse labeled "2", if the length of the pulse "1"
$t_{dn}^1$ is clearly longer then the lifetime $t_{up}^2$ of the pulse "2".
Since the time during which we observe pulse "1" is longer, therefore also
the bandwidth should be wider and the fluctuations around "1" should be
stronger then those around "2". This does not happen because RTN noise
constitutes a Markovian process without a memory. When the RTN switches to
another level, the only information available to the system is the
probability of switching back to the previous state given by the inverse of
the statistically average RTN lifetime.  As our system does not know how
long it will stay in a given state, the variance at a given level cannot
adjusts itself to the actual pulse length but only to the statistically
significant variable, i.e., to the average lifetime. Remember that RTN
lifetimes (pulse lengths) are exponentially distributed. Therefore one may
well encounter a long pulse belonging to the distribution with a short
average lifetime as well as a short pulse from the distribution
characterized by a long average lifetime. Consequently, fluctuations around
a shorter pulse with a longer average lifetime will appear stronger than
those around a longer pulse with a short average lifetime, as is the case
illustrated in Fig. \ref{last}.

\section {Conclusions}
We conclude that, at least in our experimental case, quiet and noisy
metastable states do not exist.  The difference between fluctuations around
distinct RTN levels, so clearly visible in the experimental time records, is
due only to the differences in the effective bandwidth in which one the
background noise is seen in the experiments. The bandwidth limits are
imposed by the signal sampling rate and the average lifetimes of the random
telegraph signal. The observed changes of the variance ratio with changing
current result from the current dependence of the RTN lifetimes.

We emphasize that to claim that quiet and noisy metastable states really
exist in the reality it is not enough to detect different noise variances
around different RTN levels. The proper conclusion can be drawn only after
different background fluctuations appear in a symmetric RTN waveform, for
which the  background noise bandwidth at both RTN levels is the same. In the
pioneering experiments on flux noise this was clearly not the case in the
entire temperature range investigated as it follows from the evaluations of
the average RTN lifetimes \cite{clarke}. Nevertheless, in these zero field
and zero current experiments it was not possible to change the pristine
symmetry of the observed RTN signal and to establish if the quiet and noisy
metastable states are  due to bandwidth differences or reflect a real exotic
structure of the involved two-level fluctuator. In our early communique
\cite{eucas} concerning amplitude modulated RTN voltages in thin HTSC films
we suggested the possible existence of statistical correlations between the
RTN and background noise components based on appearances of quiet and noisy
metastable states. Unfortunately, we completely disregarded the fact that
noise variances around RTN levels were actually equal for a symmetric RTN
signal. The quiet and noisy metastable states as reported in \cite{eucas}
are therefore most likely also due to the noise bandwidth limits that change
with changing bias current.

Finally, we consider why different RTN background noise levels are so rarely
observed experimentally. If the bandwidth limiting mechanism is correct, one
would expect to see quiet and noisy metastable states in all asymmetric RTN
manifestations. The answer lies in the particular functional form of the
background noise. Note that the bandwidth differences have little influence
on the variance ratio when the background noise is white. It follows from
Eq. (\ref{varint}) that for white background noise with spectral density
$C(I)$
\begin{equation}
\frac{<\sigma^2_{dn}>}{<\sigma^2_{up}>}
=\frac{C_{dn}(I)}{C_{up}(I)}\frac{\frac{1}{\Delta
t}-\frac{1}{\tau_{dn}}}{\frac{1}{\Delta t}-\frac{1}{\tau_{up}}} .
\end{equation}
Since the sampling rate $1/\Delta t$ has to satisfy $\Delta t \ll
\tau_{up},\tau_{dn}$, for the white background noise
$<\sigma^2>_{dn}/<\sigma^2>_{up}\approx 1$. Background noise with
$1/f^\alpha$-like spectrum clearly exercises a much stronger influence on
the variance ratio. In fact, only this type of background noise can give
rise to experimentally observable quiet and noisy metastable states.

\acknowledgments This work was supported by THE ISRAEL SCIENCE FOUNDATION
founded by The Academy of Sciences and Humanities, and by the Israeli
Ministry of Science and by a Polish Goverment KBN grant. The authors thank
the group of Prof. Maritato at the University of Salerno for providing the
thin film samples. Stimulating discussions with Georges Waysand and Ilan
Bloom are greatly appreciated.

\begin{figure}
\caption{ A typical telegraph noise time record demonstrating quiet (down
level) and noisy (up level) metastable states.}
\label{rtn-record} %Fig.1
\end{figure}

\begin{figure}
\caption{Current dependence of the variance ratio (left axis) and RTN
amplitude (right axis) recorded at zero applied magnetic field in current
induced dissipative state (zero cooled sample) at 77 K.
Note that $\sigma_{up}/\sigma_{dn}=1$ for  $I=I_s$=14.3 mA.}
\label{variance-ratio}%Fig.2
\end{figure}

\begin{figure}
\caption{The evolution of average RTN lifetimes with changing bias current.
Note that for the symmetrizing current $I=I_s$ the
RTN waveform is symmetric, $\tau_{up}=\tau_{dn}$.} \label{tempi}%Fig.3
\end{figure}

\begin{figure}
\caption{Amplitude histogram of a noisy RTN signal. An example of
overlapping Gaussian distributions described by Eq.
(\ref{Gauss}). } \label{gauss-fig}%Fig.4
\end{figure}

\begin{figure}
\caption{Results of the analysis of a noisy RTN signal in the time domain
according to the proposed technique. Shown is (a) original experimental time
record, (b) pure RTN component inferred from the experimental record.}
\label{example}%Fig.5
\end{figure}

\begin{figure}
\caption{Spectral densities of the total experimental record  PSD$_{exp}$,
of the pure RTN component PSD$_{rtn}$, and of the background noise
PSD$_{bckgnd}$ for currents close to $I_s$. The spectral density of a pure
RTN wave has been calculated from Machlup's formula (\ref{machlup}) using
statistically averaged values obtained from the discussed time domain
analysis. The PSD$_{bckgnd}$ is further divided into constituent components
related to $\{dn\}$ and $\{up\}$ telegraph states. The down and up
background noise components shown are evaluated from the artificial time
records containing only data points belonging to the relevant RTN state
(separate spectral components of the background noise are not normalized to
the record length). Note that all components of the background noise follow
the same $1/\sqrt{f}$ frequency dependence.}
\label{PSD}%Fig.6
\end{figure}

\begin{figure}
\caption{Variance ratio of the background noise around up and down RTN
states calculated according to Eq. (\ref{varint}), solid circles, compared
with the variance ratio obtained from the Gaussian distributions
of the experimental data, filled squares.} \label{final}%Fig.7
\end{figure}

\begin{figure}
\caption{RTN record demonstrating noisy metastable state in a short pulse.}
\label{last}%Fig.8
\end{figure}

\vfill\eject \epsfxsize=\hsize\epsfysize=\vsize\epsfbox{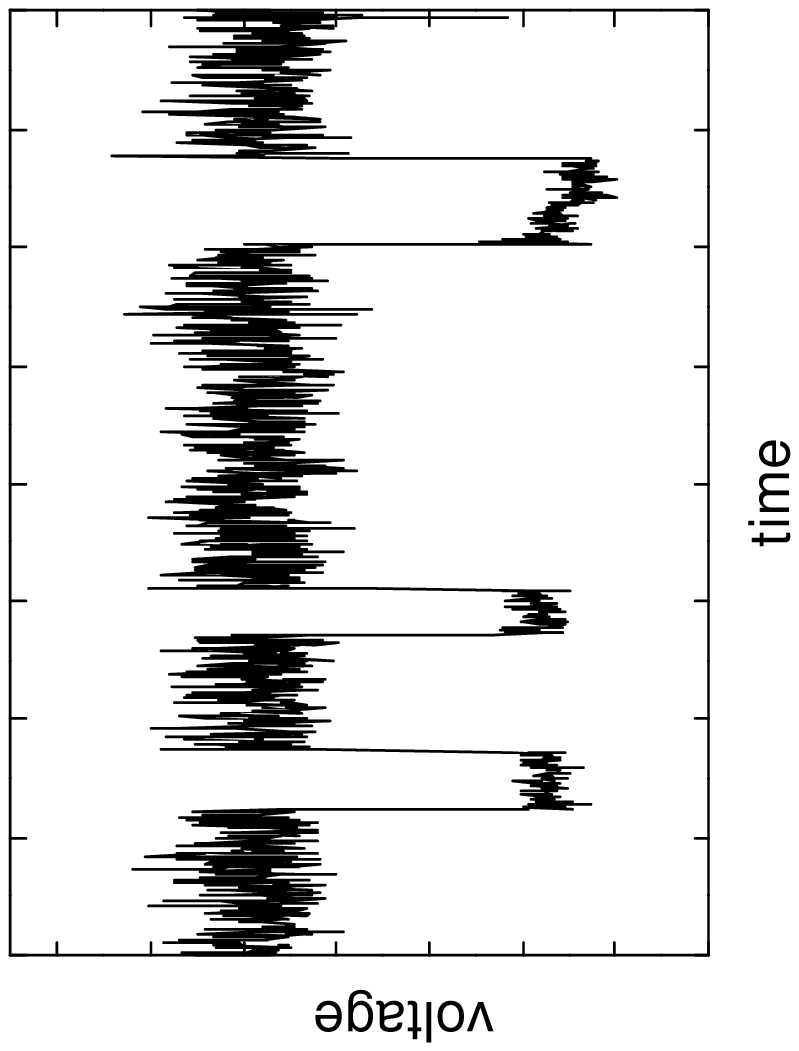}\vfill\eject
\epsfxsize=\hsize\epsfysize=\vsize\epsfbox{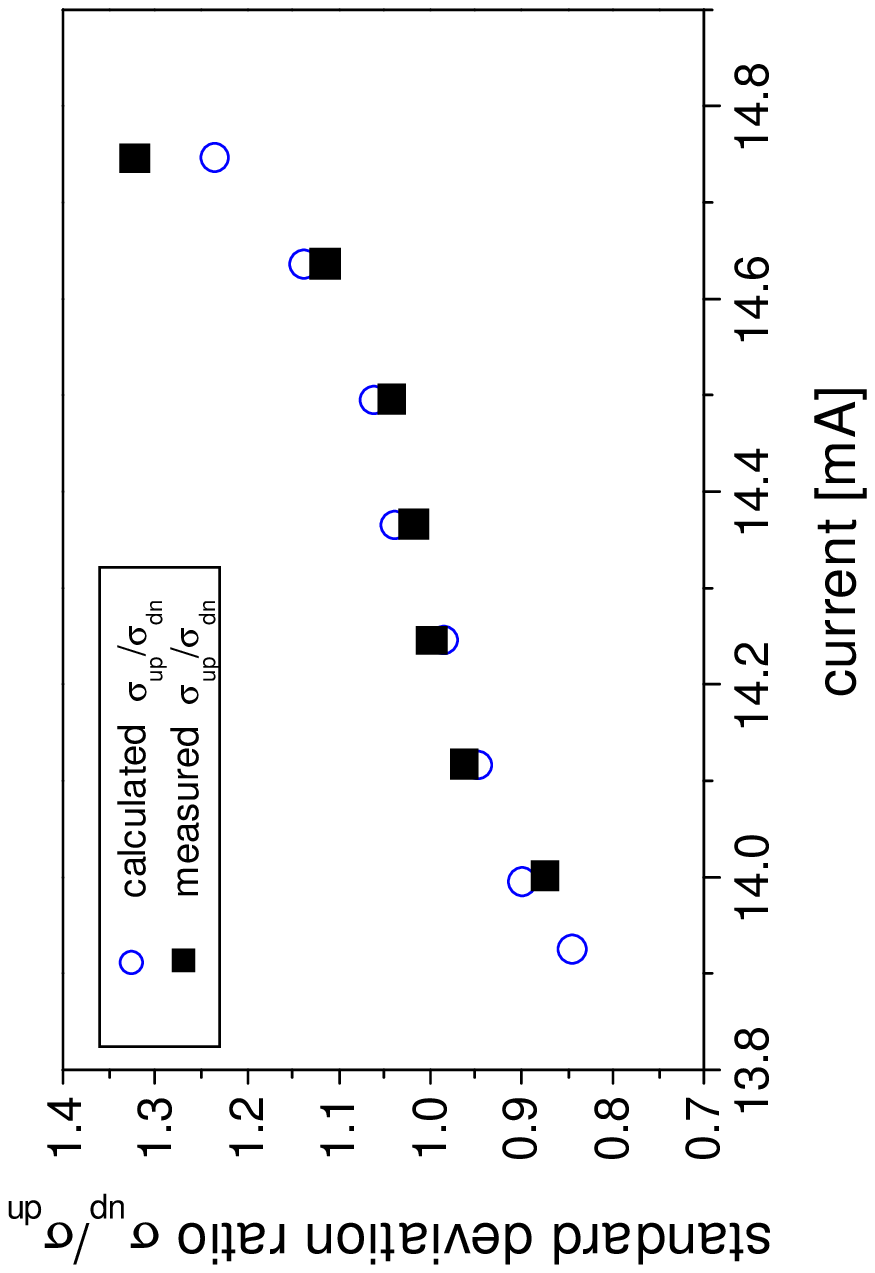}\vfill\eject
\epsfxsize=\hsize\epsfysize=\vsize\epsfbox{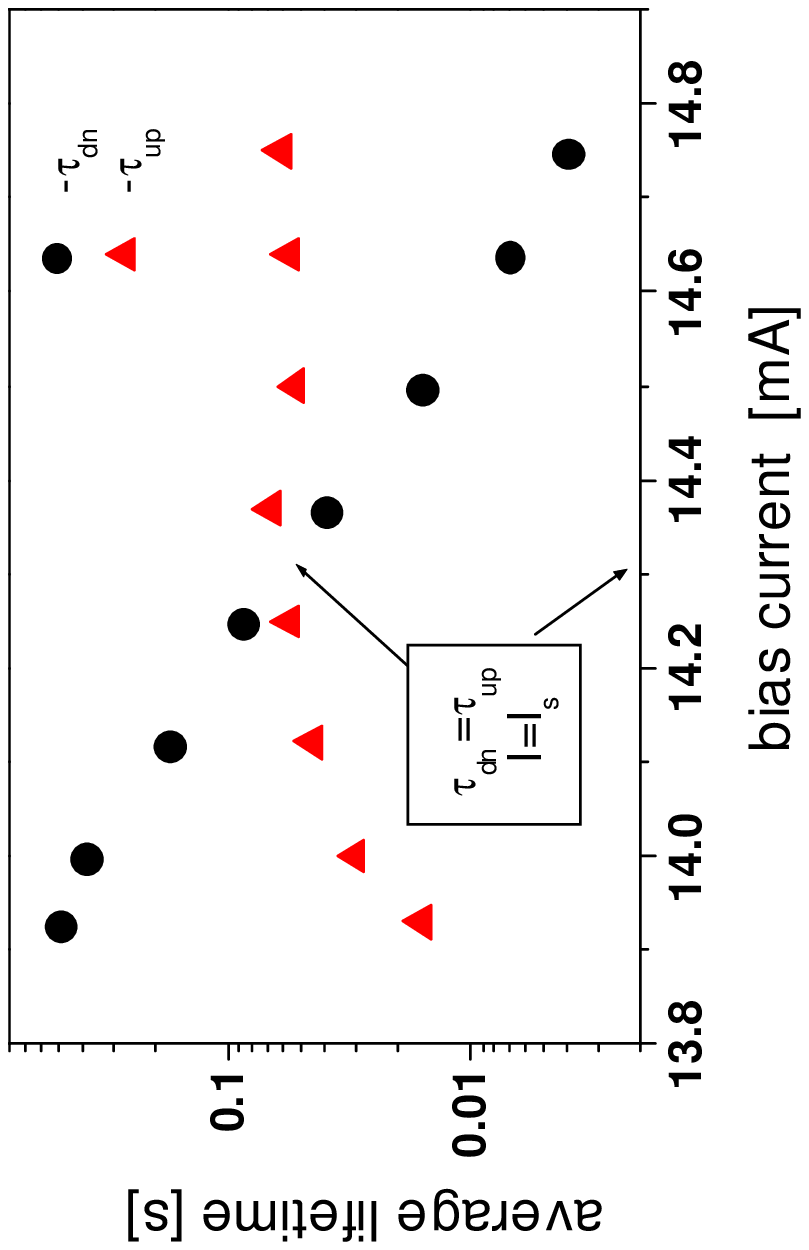}\vfill\eject
\epsfxsize=\hsize\epsfysize=\vsize\epsfbox{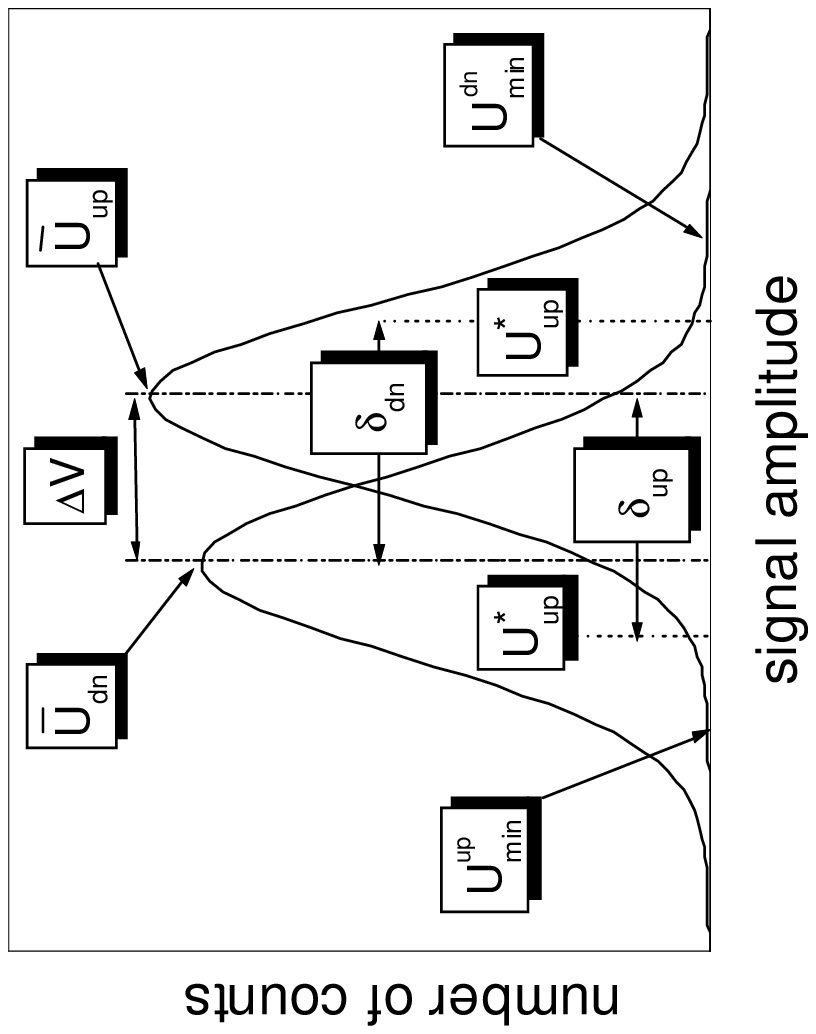}\vfill\eject
\epsfxsize=\hsize\epsfysize=\vsize\epsfbox{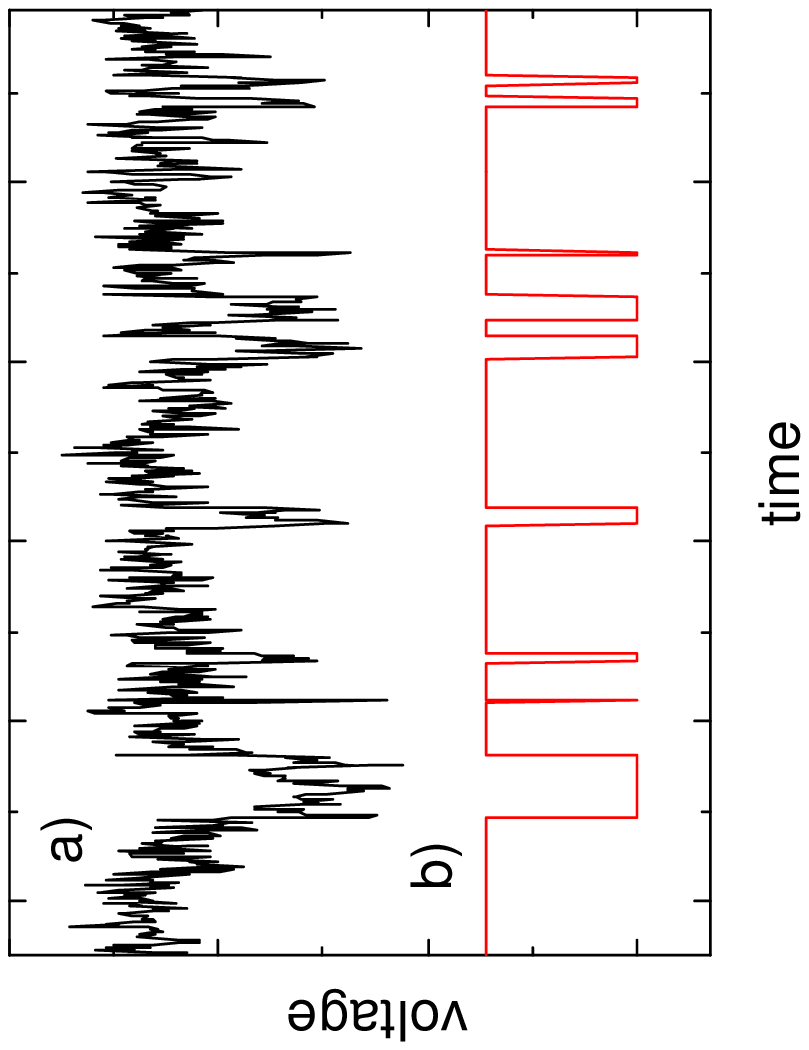}\vfill\eject
\epsfxsize=\hsize\epsfysize=\vsize\epsfbox{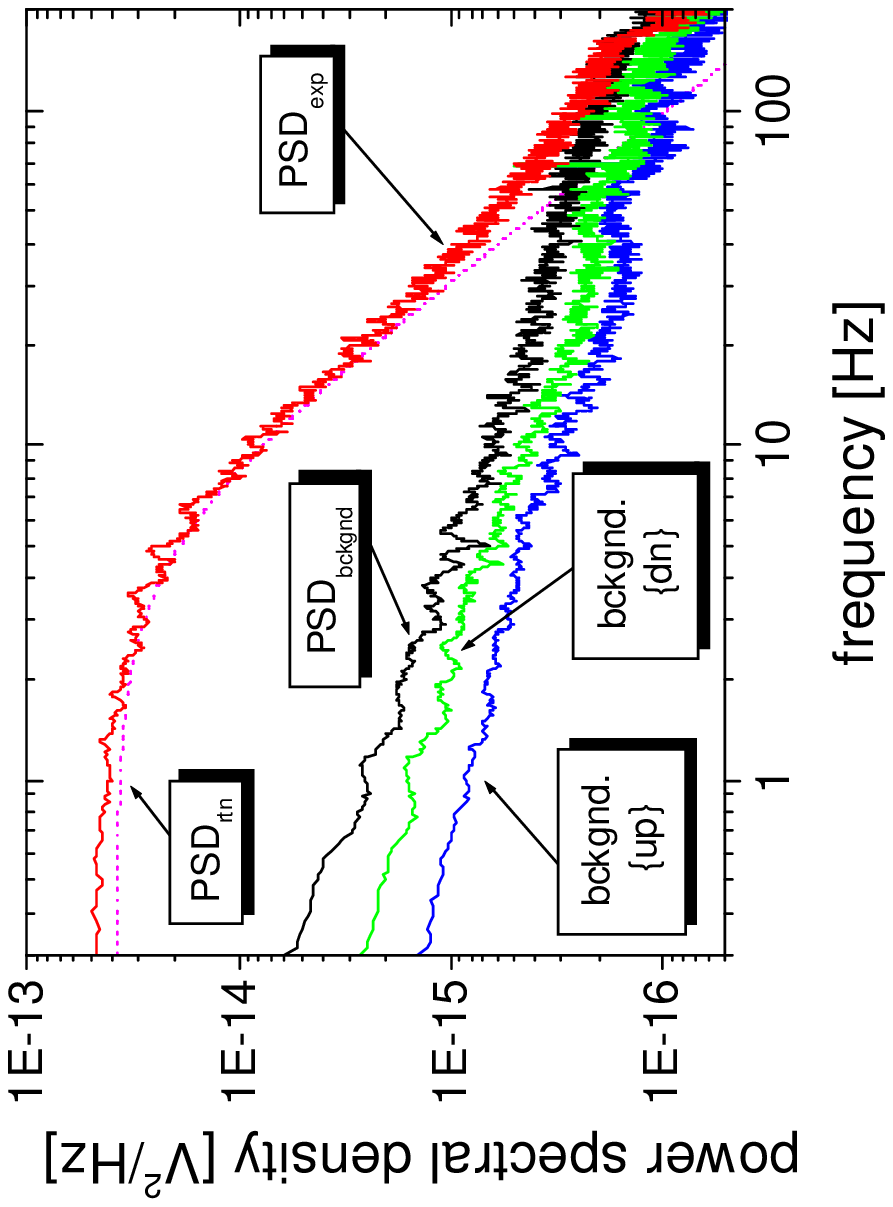}\vfill\eject
\epsfxsize=\hsize\epsfysize=\vsize\epsfbox{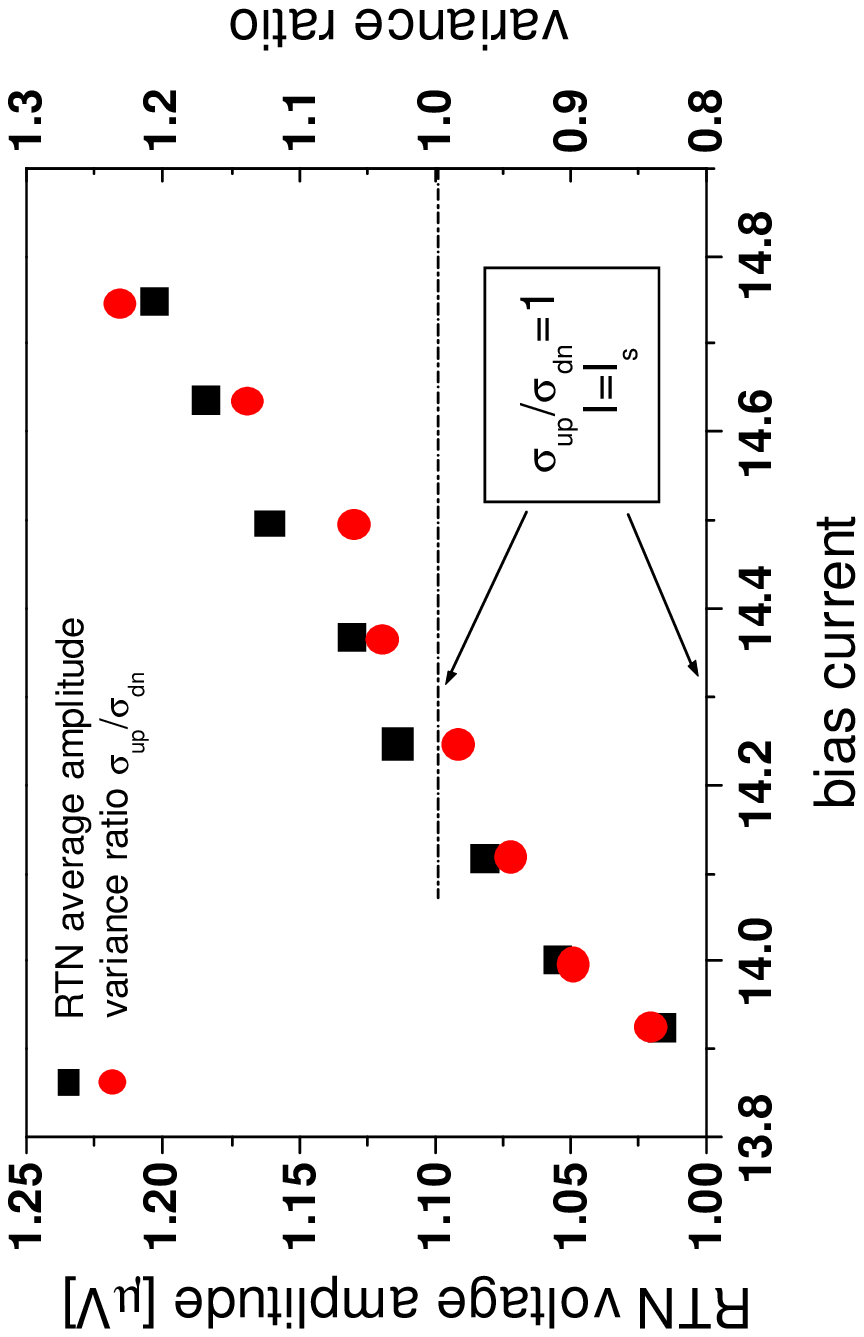}\vfill\eject
\epsfxsize=\hsize\epsfysize=\vsize\epsfbox{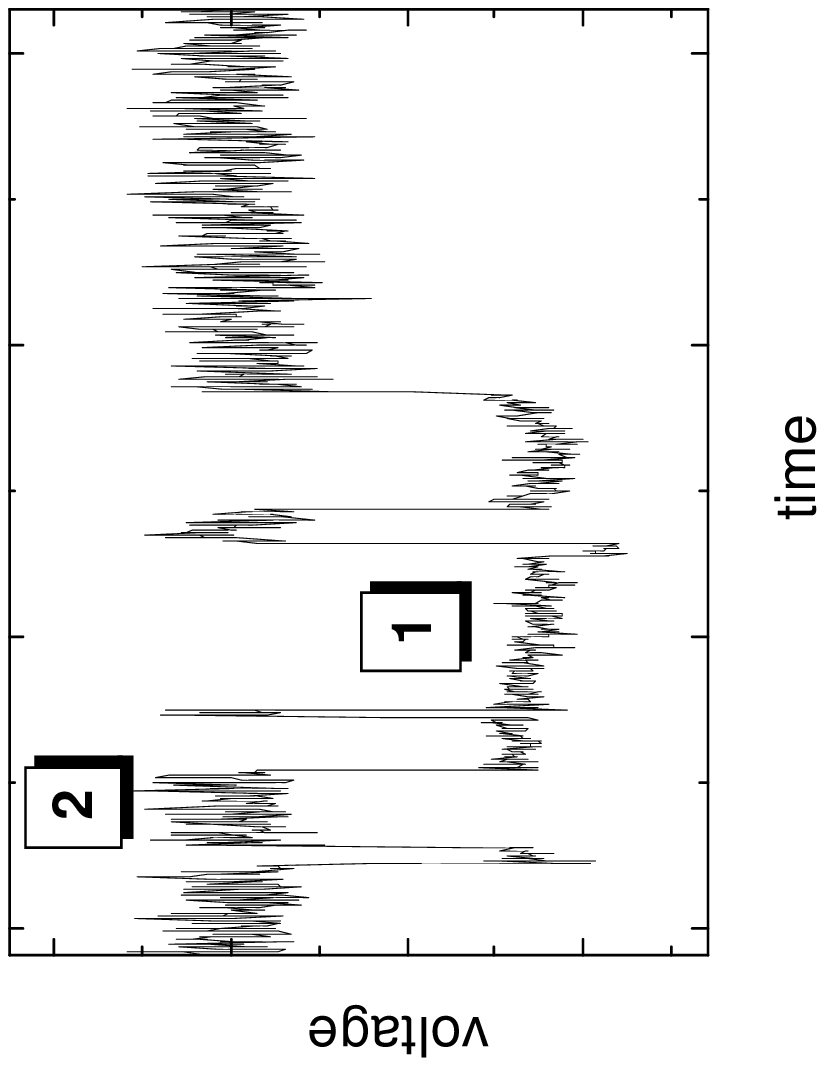}


\begin{references}
\bibitem[*]{byline} Also at Instytut Fizyki PAN, Al.Lotnik\'{o}w 32, PL
02-668 Warszawa, Poland .
\bibitem{clem}
J. R. Clem, Phys. Rep. {\bf 75}, 1 (1981)
\bibitem{ens}
B. Placais, P. Mathieu, Y. Simon, Phys. Rev. B {\bf 49}, 15813
(1994)
\bibitem{jap1f}
G. Jung and B. Savo, J. Appl. Phys. {\bf 80}, 2939 (1996)
\bibitem{koganbook}
Sh. Kogan, {\it Electronic Noise and
Fluctuations in Solids} (Cambridge University Press, Cambridge,
U.K., 1996)
\bibitem{clarke}
M. J. Ferrari, Mark Johnson, F. C. Wellstood, J. J. Kingston, T.
J. Shaw, and  John Clarke, J. Low Temp. Phys. {\bf 94}, 15 (1994)
\bibitem{clarkecorrel}
T. S. Lee, N. Missert, L. T. Sagdhal, John Clarke, J. R. Clem, K.
Char, J. N. Eckstein, D. K. Fork, L. Lombardo, A. Kapitulnik, L.
F. Shneemayer, J. V. Waszczak, and R. B. van Dover, Phys. Rev.
Lett. {\bf 74}, 2796 (1995)
\bibitem{uk}
S. A. L. Foulds, J. Smithyman, G. F. Cox, C. M. Muirhead, R. G.
Humphreys, Phys. Rev. {\bf B 55}, 9098 (1997).
\bibitem{argonne}
E. Shung, T. F. Rosenbaum, S. N. Coppersmith, G. W. Crabtree, and
W. Kwok, Phys. Rev. {\bf B 56}, R11431 (1997)
\bibitem{mmm}
R. Buder, J. Dumas, C. Escribe-Filippini, H. Guyot, Ch. J. Liu, J.
Markus, S. Revenaz, P. L. Reydet, and C. Shlenker, in {\it
"Studies of High Temperature Superconductors"}, vol. {\bf 7} ed.
A. Narlikar, (Nova Sicence Publishers, New York, p. 223).
\bibitem{jap}
G. Jung, S. Vitale, J. Konopka, and M. Bonaldi, J. Appl. Phys.
{\bf 70}, 5440 (1991 )
\bibitem{epl}
G. Jung, B. Savo, A. Vecchione, Europhys. Lett. {\bf 21}, 947
(1993)
\bibitem{isra} E. R. Nowak, N. E. Israeloff,
and A. M. Goldman Phys. Rev. {\bf B 49}, 10047 (1994)
\bibitem{kiss}
L. B. Kiss and P. Svendlindh, IEEE Trans. {\bf ED 41}, 2112 (1994)
\bibitem{prb}
M. Bonaldi, G. Jung, B. Savo, A. Vecchione, and  S. Vitale,
Phys. Rev. {\bf B 53}, 90 (1996).
\bibitem{toledo}
G. Jung, B. Savo, A. Vecchione, I. Khalfin, B. Ya. Shapiro, in
{\it "Superconductivity and Particle Detection"}, ed. T. Girard,
A. Morales, and G. Waysand, Word Scientific, Singapore 1995, p.
291
\bibitem{eucas}
G. Jung, B. Savo, C. Coccorese, in {\it Applied Superconductivity}, ed. D.
Dew Huges, Institute of Physics Publishing, Bristol 1996, p. 1003
\bibitem{upon}
G. Jung, Y. Yuzhelevski, B. Savo, C. Coccorese, V. D. Ashkenazy, B. Ya.
Shapiro, in {\it Unsolved Problems of Noise}, ed. Ch. R. Doering, L. Kiss,
and M. F. Shlesinger, World Scientific 1997, p. 285.
\bibitem{wakai}
R. T. Wakai and D. J. Van Harlingen, Phys. Rev. Lett. {\bf
58}, 1687 (1987)
\bibitem{kirton}
M. J. Kirton and M. J. Uren, Adv. Phys. {\bf 38}, 367, 1989
\bibitem{farmer}
K. R. Farmer, C. T. Rogers, R. A. Burhman, Phys. Rev. Lett. {\bf
58}, 2255 (1987)
\bibitem{others}
K. S. Ralls and R. A. Burhman, Phys. Rev. Lett. {\bf 60}, 2434 (1988)
\bibitem{mbe}
C. Attanasio, C. Coocorese, L. Maritato, M. Salluzzo, and M.
Slavato, Nuovo Cimento D {\bf 16}, 1961 (1984)
\bibitem{kleinpenning1}
T. G. M. Kleinpenning and A. H. de Kuijper, J. Appl. Phys. {\bf 63}, 49 (1988)
\bibitem{strassila}
U. J. Strasilla and M. J. O. Strutt, J. Appl. Phys. {\bf
46}, 1423 (1974)
\bibitem{kleinpenning2}
T. G. M. Kleinpenning, Physica B {\bf 164}, 331
(1990)
\bibitem{machlup}
S. Machlup, J. Appl. Phys. {\bf 25}, 341 (1954)
\end{references}
\end{document}